\documentclass[
    ,final            
  ]
  {aipproc}

\layoutstyle{6x9}

\begin{document}

\newcommand{\lapprox}{%
\mathrel{%
\setbox0=\hbox{$<$}
\raise0.6ex\copy0\kern-\wd0
\lower0.65ex\hbox{$\sim$}
}}
\newcommand{\gapprox}{%
\mathrel{%
\setbox0=\hbox{$>$}
\raise0.6ex\copy0\kern-\wd0
\lower0.65ex\hbox{$\sim$}
}}

\newcommand{\be}{\begin{equation}}
\newcommand{\ee}{\end{equation}}
\newcommand{\bea}{\begin{eqnarray}}
\newcommand{\eea}{\end{eqnarray}}

\newcommand{\gev}{\mbox{GeV}}
\newcommand{\tev}{\mbox{TeV}}
\newcommand{\SM}{\mathrm{SM}}
\newcommand{\LHT}{\mathrm{LHT}}
\newcommand{\BR}{\mathrm{BR}}


\vspace*{-1.8cm} 
\begin{flushright}
July 11, 2007 \\
HRI-P-07-07-001
\end{flushright}

\vspace*{-0.7cm} 
\title{Invisibly decaying Higgs boson in the Littlest Higgs model with
  T-parity}

%

\author{R. Srikanth Hundi}{
  address={Harish-Chandra Research Institute, Chhatnag Road, Jhusi, Allahabad
  - 211019, India}
}

\author{B.\ Mukhopadhyaya}{
   altaddress={Harish-Chandra Research Institute, Chhatnag Road, Jhusi,
  Allahabad - 211019, India}
}

\author{A.\ Nyf\/feler\footnote{Talk presented by A.N.\ at the International
  Workshop on Theoretical High Energy Physics (IWTHEP 2007), Roorkee, India,
  15-20 March 2007, to appear in the proceedings.}}{
  altaddress={Harish-Chandra Research Institute, Chhatnag Road, Jhusi,
  Allahabad - 211019, India} 
}


\begin{abstract}
We show that there are regions in the parameter space of the Littlest Higgs
model with T-parity, allowed by electroweak precision data, where the Higgs
boson can decay invisibly into a pair of heavy photons $A_H$ with a
substantial branching ratio. For a symmetry breaking scale $f$ in the range
450-600 GeV, the $\mbox{BR}(H \to A_H A_H)$ can be up to 95\% for an
intermediate mass Higgs, and from 20\% down to a few percents for a Higgs
boson of mass $200~\gev$ or above. The total decay width of the Higgs boson
can thereby be enhanced by an order of magnitude compared to the Standard
Model for Higgs masses around 130 GeV.
\end{abstract}

\maketitle


\vspace*{-0.9cm} 
\section{Introduction}

\vspace*{-0.2cm} 
Little Higgs models~\cite{LH_original} have been proposed as a
solution to the little hierarchy problem of the Standard Model (SM), i.e. the
tension between a light Higgs mass and the large scale of new physics of the
order of a few TeV from electroweak precision tests. In Little Higgs models,
the Higgs is a pseudo-Goldstone boson of a symmetry breaking that takes place
at a scale $f \sim 1~\tev$. It gets a small mass only through radiative
effects and the underlying (broken) symmetry protects the Higgs mass from
getting quadratic divergences at one loop, thereby removing the little
hierarchy problem.  In order to ease constraints from precision tests, one can
further impose a discrete symmetry, T-parity~\cite{T_parity}, which allows a
rather low scale of symmetry breaking and leads to new particles in the range
of a few hundred GeV and a natural dark matter candidate, the heavy partner of
the photon, $A_H$.

A low symmetry breaking scale $f$ raises the interesting possibility that a
heavy or even intermediate mass Higgs boson could decay `invisibly' into a
pair of stable, heavy photons $A_H$ which will not leave any traces in the
detector, thereby leading to a missing energy signature, if T-parity is
exact. Such a possibility is facilitated by the fact that the state $A_H$ can
be quite light, with $M_{A_H} \approx g^\prime f / \sqrt{5} \approx 0.15
f$. The decay $H \to A_H A_H$ in the Littlest Higgs model with T-parity (LHT)
has already been mentioned briefly in
Refs.~\cite{Asano_et_al,Chen_Tobe_Yuan}. Ref.~\cite{Asano_et_al} obtained an
invisible branching ratio of about 5\% for $M_H \sim 170~\gev$, if the
condition is imposed that the heavy photon $A_H$ should constitute all of the
dark matter in the universe. Ref.~\cite{Chen_Tobe_Yuan} studied the production
and decay of the Higgs boson in the LHT at the LHC. However, the authors of
Ref.~\cite{Chen_Tobe_Yuan} only considered scales $f \gapprox 700~\gev$ where
the branching ratio is smaller than 1\% and therefore they did not take this
decay channel into account.

Here we will summarize our work~\cite{our_paper} where we have shown that
there are regions in the parameter space of the LHT, fully compatible with
electroweak constraints, where the invisible decay $H \to A_H A_H$ can have a
substantial branching ratio. Such a large invisible decay width is unlikely
for the lightest neutral supersymmetric Higgs, at least in the minimal version
of the theory.  Therefore this invisible Higgs boson decay might help to
distinguish the LHT from the MSSM at present and future colliders.


\section{The invisible decay $H \to A_H A_H$} 

\vspace*{-0.2cm} 
The mass (squared) of the heavy T-odd photon $A_H$ is given by $M^2_{A_H} =
{{g^\prime}^2 f^2 \over 5} - {{g^\prime}^2 v_{\SM}^2 \over 4}$, neglecting
higher powers of $v^2_{\SM}/f^2$. Since the heavy photon, as the lightest
T-odd state, is stable, there are no off-shell decays $H \to A_H^* A_H^*$ and
the channel opens up only for $m_H \geq 2 M_{A_H}$.  The interaction $H
{A_H}_\mu A_H^\mu$ in the LHT is described by the Feynman rule $(-i / 2)
{g^\prime}^2 v_{\LHT} g_{\mu\nu}$. The quantity $v_{\LHT}$ is obtained from
$v_{\SM} = 246~\gev$ by inverting the relation $v_{\SM} \equiv f \sqrt{1 -
\cos (\sqrt{2} v_{\LHT} / f)}$~\cite{Chen_Tobe_Yuan}. The decay width is then
given by
\be \label{widthHtoAHAH} 
\Gamma(H \to A_H A_H) = { {g^\prime}^4 v^2_{\LHT}
\over 2048 \pi M_{A_H}^4} m_H^3 \beta_A \left( 4 - 4 a_A + 3 a_A^2 \right),
\ee
where $a_A = 1 - \beta_A^2 = 4 M_{A_H}^2 / m_H^2$. From
Eq.~(\ref{widthHtoAHAH}) we see that the partial width scales like
$1/M_{A_H}^4 \sim 1/ f^4$. 

As pointed out in Ref.~\cite{Chen_Tobe_Yuan}, the couplings of the Higgs boson
to the SM particles are subject to corrections in the LHT.  We have used the
program HDECAY~\cite{HDECAY} to calculate the partial widths of the Higgs
boson in the SM in the various channels. The corresponding decay widths in the
LHT have then been obtained in a simplified (and approximate) way by
multiplying the SM results with the appropriate correction factors. Adding the
new invisible decay mode $H \to A_H A_H$ from Eq.~(\ref{widthHtoAHAH}) leads
to the total width of the Higgs boson in the LHT.

\begin{figure}[!t]

\includegraphics[height=.26\textheight]{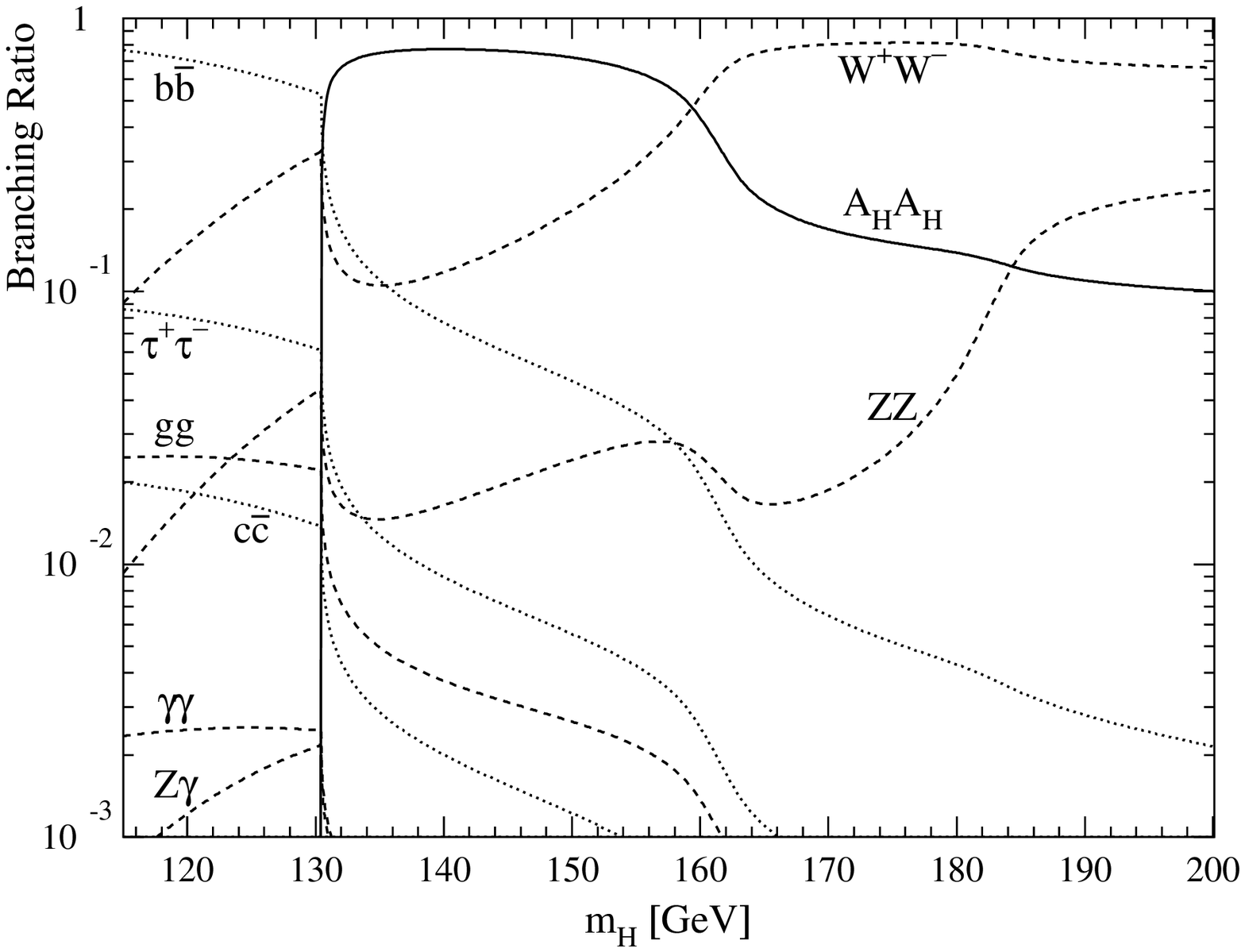}
\includegraphics[height=.26\textheight]{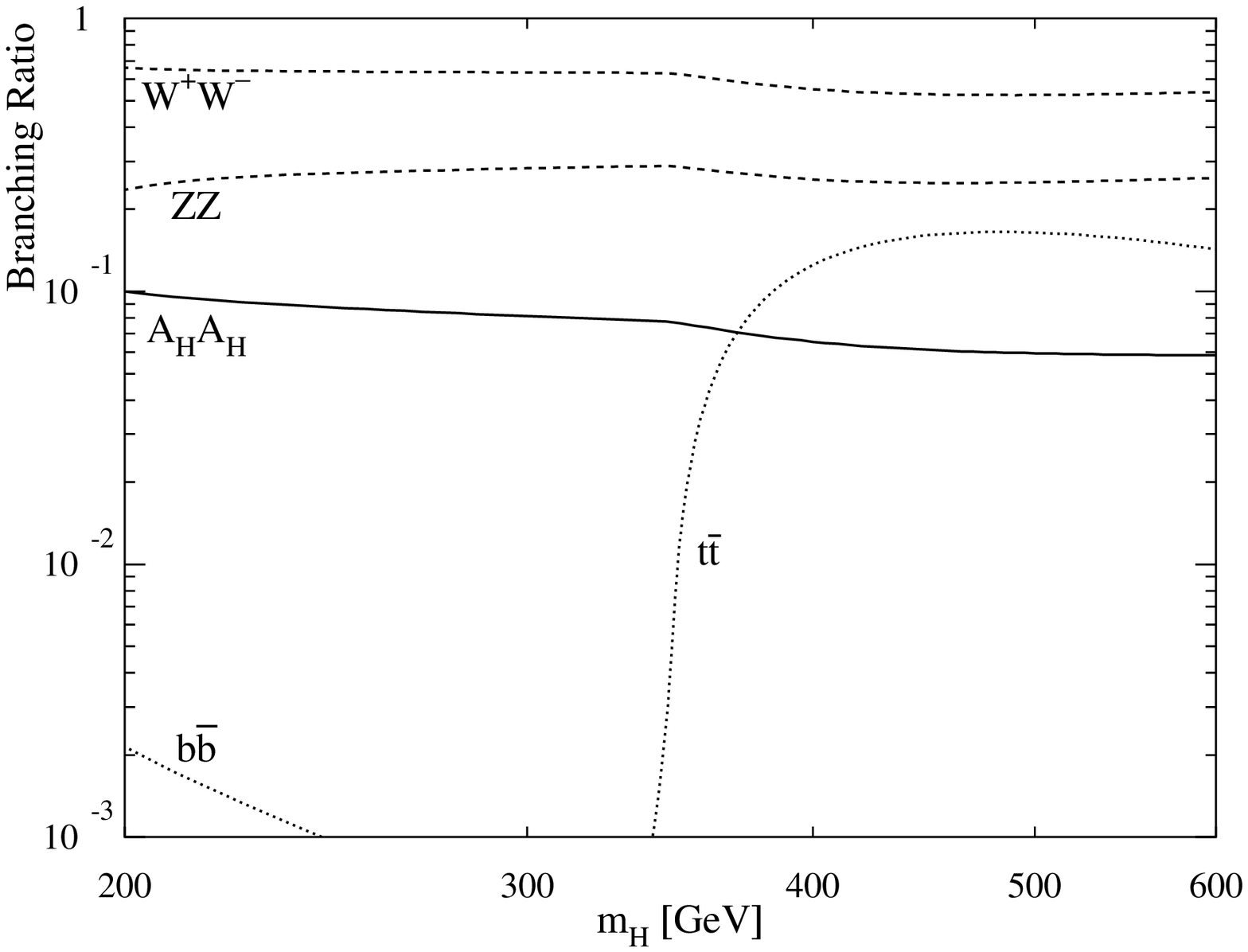}

\caption{Branching ratios in the Littlest Higgs model with T-parity for Higgs
  masses below $200~\gev$ (left panel) and above  $200~\gev$ (right panel)
  for a symmetry breaking scale $f = 500~\gev$.} 
  \label{fig:BR_LHT}

\end{figure}

In Fig.~\ref{fig:BR_LHT} we have plotted for $f=500~\mbox{GeV}$ all branching
ratios of the Higgs boson in the LHT that are larger than $10^{-3}$ in the
mass range $115~\gev < m_H < 600~\gev$. One observes, that as soon as the
decay $H \to A_H A_H$ is kinematically allowed, $m_H \geq 2 M_{A_H} =
130~\mbox{GeV}$, we get a huge invisible $\BR(H \to A_H A_H)$ of about $75\%$
in the Higgs mass range $135 - 150~\gev$. The reason is that the Higgs boson
couples to the heavy photons $A_H$ with electroweak strength $g^\prime$ which
is much larger than the Yukawa coupling to the bottom quarks. The decay width
is also larger than the off-shell (three or four-body) decay $H \to W^{(*)}
W^{*}$, unless that decay starts to grow around $m_H = 2 M_W$. At $m_H =
159~\gev$ we have $\BR(H \to A_H A_H) \approx \BR(H \to WW) = 47\%$. At $m_H =
200~(600)~\mbox{GeV}$ the invisible decay BR is still about $10~(5.5)\%$. 

Below the threshold of $130~\gev$, the same decay channels are open as in the
SM, however, $H \to gg$ is highly suppressed in the LHT, see
Ref.~\cite{Chen_Tobe_Yuan}.  Since we have taken the fermion couplings from
the ``case A'' proposed in Ref.~\cite{Chen_Tobe_Yuan}, which differ not much
from their SM values, there is no large enhancement of the $H\to \gamma\gamma$
mode as observed in that reference for the ``case B''. As soon as the decay $H
\to A_H A_H$ is possible, all other branching ratios drop down considerably.

Note that we have not taken into account the off shell-decays $H \to W_H^{*}
W_H^{*}$ and $H \to Z_H^{*} Z_H^{*}$ for Higgs masses larger than $M_{W_H} =
M_{Z_H} = 317~\gev$ for $f = 500~\gev$. We expect the corresponding branching
ratios to be small below the two-particle threshold.

\begin{figure}[!t]

\includegraphics[height=.26\textheight]{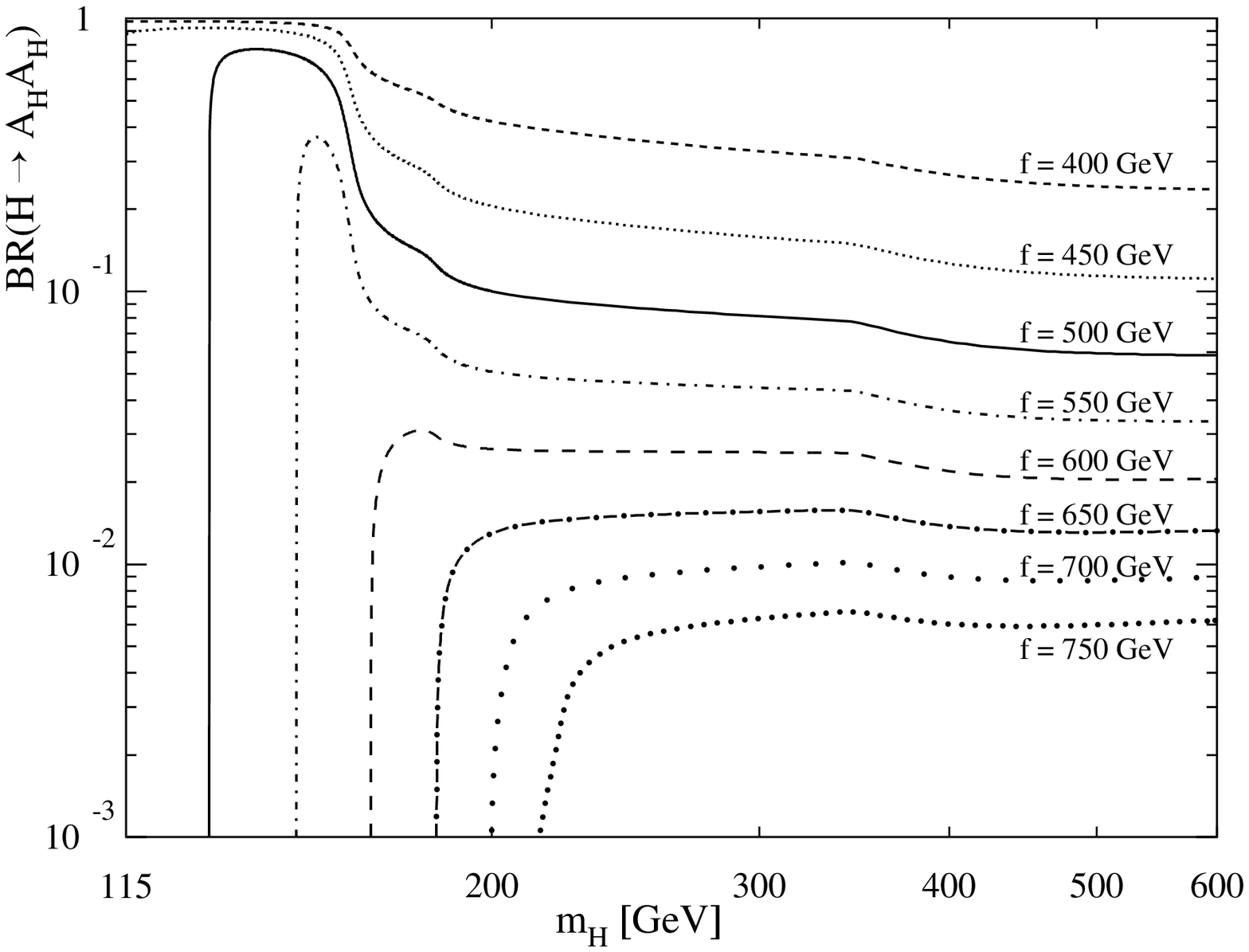}
\includegraphics[height=.26\textheight]{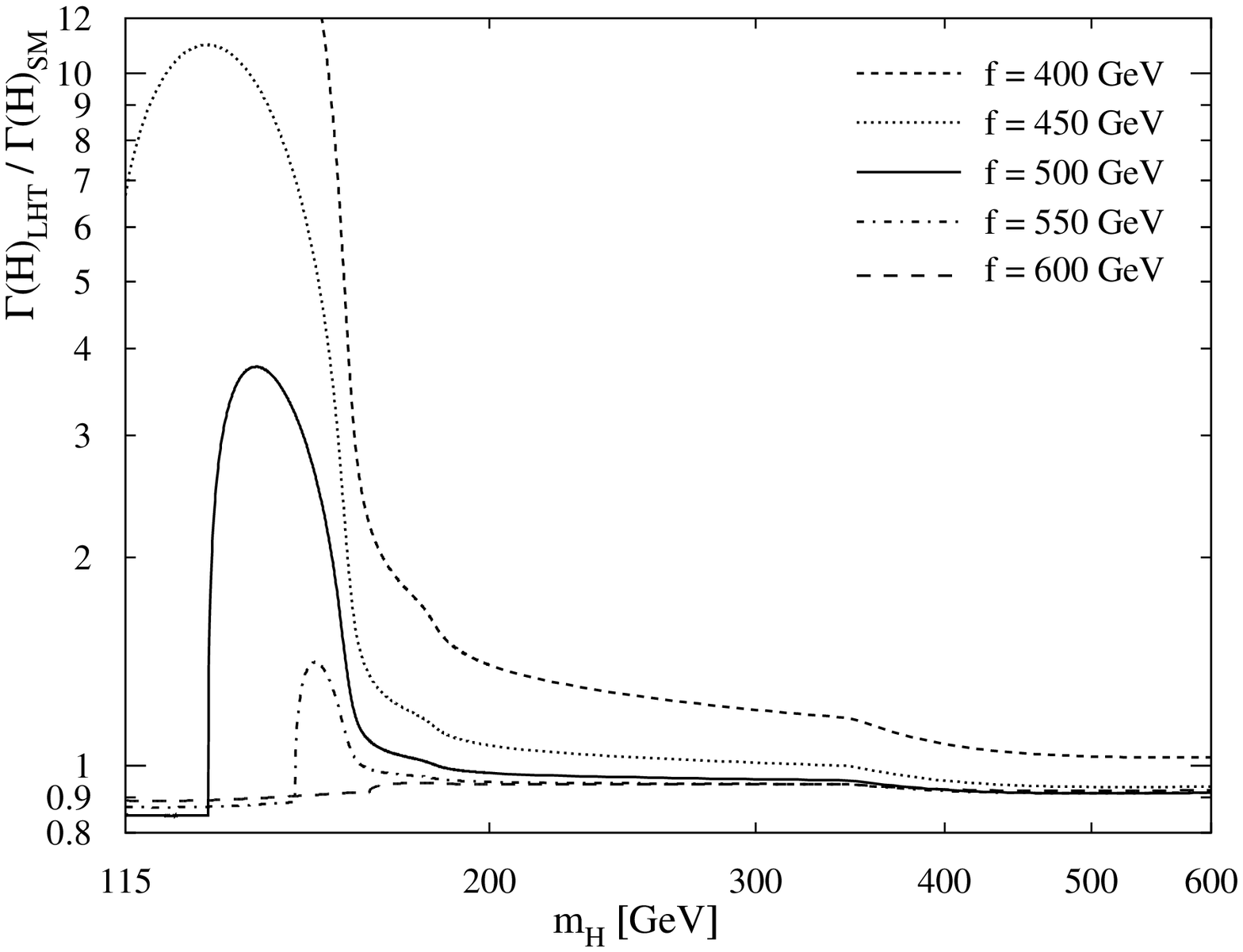}

\caption{Left panel: Branching ratio for the invisible decay $H \to A_H A_H$
  in the Littlest Higgs model with T-parity for several values of the symmetry
  breaking scale~$f$. Right panel: Ratio of the total decay width of the
  Higgs boson in the LHT, $\Gamma(H)_{\LHT}$, to the total decay width in the
  SM, $\Gamma(H)_{\SM}$, for different values of $f$. The curve for $f =
  400~\gev$ peaks at a value of about 35 for $m_H \approx 126~\gev$.}
\label{fig:BR_AHAH_Gamma_ratio}

\end{figure}

In Fig.~\ref{fig:BR_AHAH_Gamma_ratio}~(left panel) we show the invisible
branching ratio $H \to A_H A_H$ as a function of the Higgs mass for different
values of $f$ in the range $400 - 750~\mbox{GeV}$. For $f =
400~(450)~\mbox{GeV}$ the branching ratio can be as large as 98~(93)\% for
Higgs masses below about $150~\mbox{GeV}$. Although values of $f =
400~\mbox{GeV}$ are allowed by the precision data, the calculation cannot be
completely trusted there. For such low values of $f$, higher derivative terms
in the low-energy expansion in the LHT model, generated at the scale $\Lambda
\sim 4\pi f$, should be taken into account.
In this sense $f \simeq 400~\gev$ marks the lower end of the parameter
space where the underlying framework is reliable.

For $f = 600~\gev$, the invisible branching ratio is $2-3\%$ for $m_H \gapprox
169~\gev$, whereas it drops below 1\% for $f \geq 700~\mbox{GeV}$. Since the
$A_H A_H$-threshold for $f= 600~\gev$ is about $167~\mbox{GeV}$, i.e.\ above
the $WW$-threshold, the on-shell decays into $WW$ and later into $ZZ$
overwhelm the invisible decay $H\to A_H A_H$.  

Figure~\ref{fig:BR_AHAH_Gamma_ratio}~(right panel) shows the ratio of the
total decay width of the Higgs boson in the LHT, $\Gamma(H)_{\LHT}$, to the
total width in the SM, $\Gamma(H)_{\SM}$, as a function of the Higgs mass for
a subset of values of $f$ used in Fig.~\ref{fig:BR_AHAH_Gamma_ratio}~(left
panel). In contrast to earlier studies~\cite{Higgs_decays_LH,Chen_Tobe_Yuan}
which always observed a reduction of the total decay width of the Higgs boson
in the Littlest Higgs model and in the LHT compared to the SM, we get a
potentially huge enhancement of the decay width for values $f \leq
550~\mbox{GeV}$.  For $f = (400, 450, 500, 550)~\gev$, the maximal enhancement
factors of $(34.8, 11.0, 3.77, 1.41)$ that can be seen in
Fig.~\ref{fig:BR_AHAH_Gamma_ratio}~(right panel) correspond to
$\Gamma(H)_{\LHT} = (140, 51, 30, 33)~\mbox{MeV}$ at $m_H = (125.8, 130.2,
140.5, 153.4)~\gev$. Note, however, that the width of the Higgs boson in the
SM is very small for Higgs masses below the $WW$-threshold.  Only for $f \geq
600~\gev$ we obtain a reduction of the total width for the whole range of
Higgs masses $115~\gev < m_H < 600~\gev$. The ratio $\Gamma(H)_{\LHT} /
\Gamma(H)_{\SM}$ varies between $0.89$ and $0.95$ for values of $f = 600 -
750~\gev$.

We would like to stress that there are regions in parameter space where values
of $115~\gev < m_H < 650~\gev$ and $400~\gev < f < 700~\gev$ are allowed
by the electroweak data at 95\% confidence level, see Ref.~\cite{our_paper}
for details of the electroweak fit.


\section{Conclusions}

A substantial branching ratio into the invisible channel not only makes the
Higgs boson a rather interesting object but also helps in associating it with
some specific types of non-standard physics.  For example, in supersymmetric
theories, the lightest neutral scalar can in principle decay into two lightest
neutralinos, making it invisible.  However, the branching ratio of such a
decay is usually not very high, and is rather restricted in the regions of the
parameter space allowed by LEP data, at least in those versions of the theory
not too far from the minimal model. In the LHT, however, the invisible
branching ratio can not only be appreciable but also may correspond to a Higgs
boson that is heavier than what is allowed in a minimal supersymmetric
framework. Thus this region of the parameter space may provide a test to
distinguish the LHT from supersymmetry.



\end{document}